\documentclass[
aps, prb,showpacs,superscriptaddress,numerical,amsmath,amssymb,floatfix,
reprint]{revtex4-2}

\usepackage[
pdffitwindow=true,
colorlinks=true,
frenchlinks=false,
linkcolor=blue,
anchorcolor=blue,
citecolor=blue,
filecolor=blue,
urlcolor=blue,
bookmarks=true,
bookmarksopen=true,
bookmarksnumbered=true,
bookmarksopenlevel=1,
plainpages=false,
pdfpagelayout=TwoPageLeft,
pdfpagelabels=true,
breaklinks
]{hyperref}
\usepackage[per-mode=symbol,separate-uncertainty]{siunitx}
\usepackage{graphicx}
\usepackage{dcolumn}
\usepackage{url}
\usepackage{color}
\usepackage{graphicx,wrapfig,lipsum}
\usepackage{bm}
\usepackage[english]{babel}

\usepackage{color}
\usepackage{ulem}
\usepackage{float}
\usepackage{lipsum}
\bibpunct{[}{]}{,}{n}{}{}
\usepackage{chemformula}

\begin{document}
	\title{
	Temperature dependence of the  magnon-phonon interaction in high overtone bulk acoustic resonator-ferromagnetic thin film hybrids}

	\author{M.~M\"uller}
	\email{manuel.mueller@wmi.badw.de}
	\affiliation{Walther-Mei{\ss}ner-Institut, Bayerische Akademie der Wissenschaften, 85748 Garching, Germany}
	\affiliation{TUM School of Natural Sciences, Technical University of Munich, 85748 Garching, Germany}

	\author{J.~Weber}
	\affiliation{Walther-Mei{\ss}ner-Institut, Bayerische Akademie der Wissenschaften, 85748 Garching, Germany}
	\affiliation{TUM School of Natural Sciences, Technical University of Munich, 85748 Garching, Germany}

	\author{S.T.B.~Goennenwein}
	\affiliation{Department of Physics, University of Konstanz, 78457 Konstanz, Germany}
	\author{S.~Viola Kusminskiy}
	\affiliation{Institute for Theoretical Solid State Physics, RWTH Aachen, 52074 Aachen, Germany}
	\affiliation{Max Planck Institute for the Science of Light, 91058 Erlangen, Germany}
	
	\author{R.~Gross}
	\affiliation{Walther-Mei{\ss}ner-Institut, Bayerische Akademie der Wissenschaften, 85748 Garching, Germany}
	\affiliation{TUM School of Natural Sciences, Technical University of Munich, 85748 Garching, Germany}

	\affiliation{Munich Center for Quantum Science and Technology (MCQST), 80799 Munich, Germany}
	
	\author{M.~Althammer}
	\email{matthias.althammer@wmi.badw.de}
	\affiliation{Walther-Mei{\ss}ner-Institut, Bayerische Akademie der Wissenschaften, 85748 Garching, Germany}
	\affiliation{TUM School of Natural Sciences, Technical University of Munich, 85748 Garching, Germany}

	\author{H.~Huebl}
	\email{hans.huebl@wmi.badw.de}
	\affiliation{Walther-Mei{\ss}ner-Institut, Bayerische Akademie der Wissenschaften, 85748 Garching, Germany}
	\affiliation{TUM School of Natural Sciences, Technical University of Munich, 85748 Garching, Germany}

	\affiliation{Munich Center for Quantum Science and Technology (MCQST), 80799 Munich, Germany}
	\begin{abstract}
Tailored magnon-phonon hybrid systems, where high overtone bulk acoustic resonators couple resonantly to the magnonic mode of a ferromagnetic thin film, are considered optimal for the creation of acoustic phonons with a defined circular polarization. This class of devices is therefore  ideal for the investigation of phonon propagation properties and assessing their capacity to transport angular momentum in the classical and potentially even in the quantum regime. Here, we study the coupling between the magnons in a ferromagnetic \ch{Co25Fe75} thin film and the transverse acoustic phonons in a bulk acoustic wave resonators formed by the sapphire substrate onto which the film is deposited. Using broadband ferromagnetic resonance experiments as a function of temperature, we investigate the strength of the coherent magnon-phonon interaction and the individual damping rates of the magnons and phonons participating in the process.  This demonstrates that this coupled magnon-phonon system can reach a cooperativity $C\approx 1$ at cryogenic temperatures. Our experiments also showcase the potential of strongly coupled magnon-phonon systems for strain sensing applications.
	\end{abstract}
	\maketitle
\section{Introduction}
Spin-orbit coupling and its consequences like a finite magnetoelastic coupling (MEC) can transduce fundamental excitations of solid state systems between different forms. For magnetic excitations, the conversion of excitations of the spin system (magnons) to those of the lattice (phonons) typically manifest as damping of the magnetization dynamics  \cite{Spencer1958, Widom2010, Vittoria2010, Rossi2005, Strongin1976, Kobayashi1973,Klingler:2017bu}. On the other hand, the same MEC allows one to coherently excite magnetization dynamics by elastic waves\cite{Weiler2011, Weiler2012, Gowtham2016, Ku2020, Kuss2021, Kuss2021a, Kikkawa2016, Hashimoto2018a, Zhang2020a, Hatanaka2022} to control the magnetization direction by elastic strain \cite{Weiler2009, Uchida2011, Weiler2011, Weiler2012, Thevenard2014,Thevenard2013}, or to generate perpendicular magnetic anisotropy\cite{Weiler2011, Dreher2012,Qurat-Ul-Ain2020,Bi2010,An2016a}. Thus, the magnetoelastic interaction with its potential to couple magnons and phonons is considered a route towards strong magnon-phonon coupling which shows in form of hybridized excitations\cite{Yahiro2020, Sukhanov2019, Hayashi2018, Brataas2020, Kikkawa2016}. Recently, the controlled magnon-phonon coupling in geometrically tailored systems or multi-layer systems was discussed to study and control magnetization damping via phonons\cite{Streib2018,Sato2021}, the excitation of helical phonons\cite{, An2020, An2022}, the study of strong coupling in magnon-phonon hybrids\cite{An2023, Schlitz2022, An2023}, and applications in quantum sensing and transduction\cite{Sharma2022a, Bittencourt2022, Wachter2021, Potts2021,Keshtgar2014, Flebus2017, Sharma2022,Graf2021,Engelhardt2022a,Hatanaka2022}. Several seminal experiments explore bulk acoustic wave (BAW) resonators based on gadolinium gallium garnet (\ch{Gd3Ga5O12}, GGG) substrates carrying a heteroepitaxial ferrimagnetic yttrium iron garnet (\ch{Y3Fe5O12}, YIG) thin film \cite{An2020, An2022, Schlitz2022}. The choice for this particular material system is mostly motivated by the exceptional magnetic damping properties of YIG \cite{Klingler:2017bu, Maier-Flaig2017b, Jermain2017, Spencer1959}. However, YIG requires substrate materials allowing for heteroepitaxial growth. Moreover, YIG shows enhanced magnetization damping at cryogenic temperatures \cite{Maier-Flaig2017b,Jermain2017, Lachance-Quirion2019a}, where reduced acoustic losses \cite{Transducers1964, Landau1937} in combination with an increased MEC \cite{Barangi2015} potentially offer improved cooperativities $C$. Complementary, ferromagnetic metallic thin films with ultra-low magnetic damping properties, such as \ch{Co25Fe75}(CoFe), allow for the creation of magnon-phonon hybrids with a broad selection of substrate materials acting as the BAW resonator\cite{Selfcife2022}. Here, we report on the temperature dependence of the acoustic and magnetic damping as well as the magnetoelastic coupling rate in ferromagnetic metal/BAW resonator substrate bilayer systems.

In our experiments, we study the temperature and frequency evolution of the magnetic relaxation rate $\kappa_\mathrm{s}$, the elastic relaxation rate $\eta_{\mathrm{a}}$, as well as the effective magnetoelastic coupling rate $g_{\mathrm{eff}}$ for magnetic CoFe thin films deposited on \ch{Al2O3} BAW resonators. We find that the results for both $\kappa_\mathrm{s}$ and $\eta_{\mathrm{a}}$ can be well understood within the models of Gilbert damping in $3d$ transition metals\cite{Gilbert2004, Zhao2016, Schoen2016,Schoen2017,Schoen2017a} and Landau-Rumer theory\cite{Landau1937, Transducers1964}, respectively. In addition, we investigate the temperature dependence of the magnon-phonon coupling rate $g_{\mathrm{eff}}$ and the resulting cooperativity $C$ to identify optimal operation conditions for mode hybridization\cite{An2020,An2022}. We also assess the potential of magnon-phonon hybrids for strain sensing. 

\section{Experimental details}
\begin{figure}[tbh]	
		\centering
		\includegraphics[width=1.0\columnwidth, clip]{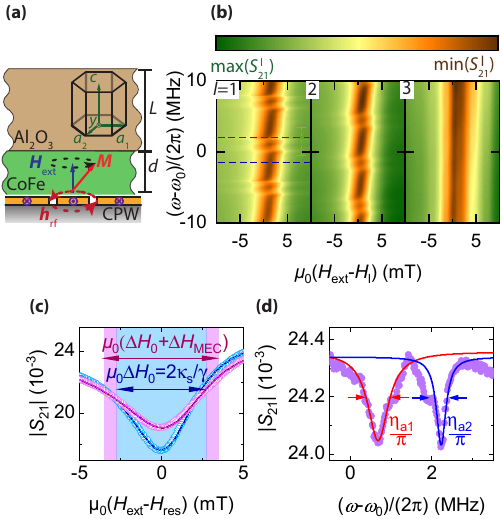}
		\caption{(a) Sketch of the experimental configuration with the sample mounted face-to-face on a coplanar waveguide. (b) CPW microwave transmission $|S_{21}|$ plotted as a function of frequency $\omega$ around $\omega_0/(2\pi)=18$ GHz and external magnetic field $H_{\mathrm{ext}}$ close to the FMR resonance field $H_l$ for various temperatures indexed by $l$ (see Tab.\,\ref{Tab:1}). (c) Fixed frequency $|S_{21}|$ showing the FMR microwave absorption on resonance (magenta) and off resonance (blue) with the ${n}^{\mathrm{th}}$ bulk elastic resonance mode. The respective frequencies are indicated by dashed lines in panel (b). (d) $|S_{21}|$ plotted versus frequency for a fixed magnetic field detuned from the FMR resonance field (see vertical violet dashed line in panel (b) $l=1$). The plot shows the absorption signatures of two elastic modes (violet symbols) and allows for the extraction of the acoustic relaxation rates $\eta_{\mathrm{a}1/2}$ by a Lorentzian fit (red and blue lines). }
		\label{Fig:1}
	\end{figure}

Our ferromagnet/insulator system comprises a Pt(3\,nm)/Cu(3\,nm)/CoFe(30\,nm)/ Cu(3\,nm)/Ta(\,3nm) multilayer stack deposited via dc magnetron sputtering on a $L=510$\,$\mu$m thick sapphire [$\mathrm{Al_{2}O_{3}}$ (0001)] substrate, which is polished on both sides. The Pt(3\,nm)/Cu(3\,nm)-seed layer ensures optimal magnetization damping properties of the CoFe film, while enabling elastic coupling to the standing wave phonons in the substrate\cite{Flacke2019, Edwards2019, Selfcife2022}. The Cu(3\,nm)/Ta(3\,nm) top layer serves for corrosion protection. To analyze the magnon-phonon coupling, we perform broadband ferromagnetic resonance (FMR) experiments in a cryogenic environment. In particular, the sample is mounted face-down onto a coplanar waveguide (CPW), such that the Ta top layer is facing the CPW. We measure the complex microwave transmission parameter $S_{21}$ of the CPW as function of frequency $\omega$ and the static applied magnetic field $H_{\mathrm{ext}}$ using a vector network analyzer (VNA). The external static magnetic field is applied along the out-of-plane direction resulting in the generation of standing circularly polarized elastic shear waves in the ferromagnet due to magnetoelastic coupling between the spin system and the lattice\cite{An2020, Schlitz2022, Selfcife2022}. A sketch of the measurement setup and geometry is presented in Fig.\,\ref{Fig:1}(a).
\section{Experimental results}
In Fig.\,\ref{Fig:1}(b), we show exemplary transmission data $|S_{21}|$ for $T=5\,$, $100\,$, and $300\,$K (corresponding to $l=1,2,3$) as function of applied magnetic field $H_{\mathrm{ext}}$ around a center frequency of $\omega_0/(2\pi)=18$\,GHz. The parameters of the sub-panels are listed in Tab.\,\ref{Tab:1}. Note that the small absolute values of $|S_{21}|$ are caused by attenuation in the microwave lines. We observe the characteristic microwave response of the Kittel mode in brown color, which features a distinct pattern being periodic along the frequency axis (visible in panel (b) for $l=1$ and $2$). The measured transmission data clearly indicate the interaction of the homogeneous FMR mode with the equally spaced high overtone elastic modes of the BAW resonator (cf. Ref.\,\cite{An2020, An2022, An2023, Schlitz2022, Selfcife2022, Chu2017}). At low temperatures, we observe a double peak signature with a periodicity of $\omega_{\mathrm{FSR}}/(2\pi)\approx6.04$\,MHz. This value agrees well with the frequency spacing $\omega_{\mathrm{FSR}}=2\pi f_{\mathrm{FSR}}$ of the standing transverse acoustic waves in the BAW resonator if we assume a total thickness $d+L$ of the resonator \footnote{$L=510\,\mu$m, $d=30\,$nm, $v_{\mathrm{t}}=6.17$\,km/s and $\tilde{v}_{\mathrm{t}}=3.17$\,km/s\cite{Selfcife2022}.}. Here, $v_{\mathrm{t}}$ and $\tilde{v}_\mathrm{t}$ are the velocities of transverse elastic shear waves in the \ch{Al2O3} substrate of thickness $L$ and CoFe film of thickness $d$, respectively. The detection of two neighboring acoustic wave resonances with a small frequency splitting of $\Delta \omega/(2\pi)\approx1.40$\,MHz is attributed to the fact that there is a small difference of the velocities of the fast and slow transverse modes $v_{\mathrm{ft}}$ and $v_{\mathrm{st}}$ ($v_{\mathrm{ft}}-v_{\mathrm{st}}\simeq$1\,m/s) due to a small but finite angle between the $c$-axis of sapphire and the propagation direction of the shear waves, which is parallel to the surface normal\cite{Selfcife2022}. For higher temperatures, e.g. $T=100\,$K ($l=2$), the phonon resonance double features shift to lower frequencies compared to panel $l=1$ due to the thermal expansion of the sample and the temperature dependence of the elastic parameters. They also become less pronounced due to an increase in acoustic damping, which in addition results in a broadening of the acoustic wave resonances. At $T=300\,$K ($l=3$), the interaction between magnons and the bulk acoustic resonator modes is no longer directly visible in the raw $|S_{21}|$-spectra. 
	\begin{table}[tbh]	
		\centering
		\begin{tabular}{|c|c|c|c|}
			\hline	
			$l$	&	1& 2 &3 \\ \hline
			$T$\,(K)	&	5 & 100 & 300 \\ \hline
			$\mu_0H_{\mathrm{res}}$\,(T)	&	3.01 & 3.00 & 2.94 \\ \hline
			$\mathrm{min}(|S_{21}^\mathrm{l}|)$\,($10^{-3}$) 	&	16 & 14 & 14 \\ \hline
			$\mathrm{max}(|S_{21}^\mathrm{l}|)$\,($10^{-3}$) 	&	25 & 21 &18\\ \hline
		\end{tabular}
		\caption{Temperature $T$, resonance field $\mu_0H_{\mathrm{res}}$, as well as minimum and maximum $S_{21}$-parameter $\mathrm{min}(|S_{21}^\mathrm{l}|)$ and $\mathrm{max}(|S_{21}^\mathrm{l}|)$ of the sub-panels ($l=1,2,3$) presented in Fig.\,\ref{Fig:1}\,(b). }
		\label{Tab:1}
	\end{table}
	
To analyze this data further, we quantify the FMR linewidth $\Delta H$ from fixed frequency cuts in our data. Specifically, we fit $|S_{21}(H_{\mathrm{ext}})|$ to a modified form of the Polder susceptibility \cite{Polder1949, Dreher2012} represented by Eq.\,\eqref{formula: s21-real} to extract the (half width at half maximum) linewidth $\Delta H$  (see Appendix\,\ref{AppendixA}).
The magnon-phonon coupling signature displayed in Fig.\,\ref{Fig:1}(b) presents itself as modification of the FMR linewidth, when displayed as fixed-frequency cuts. Figure \ref{Fig:1}(c) shows these cuts for the case when the FMR is resonant (magenta symbols) with a BAW resonator mode and when they are decoupled (blue symbols). Apparently, we observe an increase of the FMR linewidth when magnons and the excitations of the BAW resonator interact. We quantify these FMR linewidth modification by $\Delta H_{\mathrm{MEC}}=\Delta H-\Delta H_0$,  referencing the observed linewidth to the FMR linewidth for the decoupled case $\Delta H_0$.
Note that for quantifying magnon-phonon hybrids, the intrinsic magnon relaxation rate $\kappa_{\mathrm{s}}$ is a relevant parameter, which we determine  from the FMR linewidth $\Delta H_0$.

We find $\kappa_{\mathrm{s}}/(2\pi)=\gamma\mu_0\Delta H_0/(4\pi)= (69.0\pm0.1)$\,MHz for $T=5\,$K and 18\,GHz, which agrees well with Refs.\,\cite{Flacke2019, Edwards2019,Selfcife2022, Schoen2016}. 

The acoustic damping rates $\eta_{\mathrm{a1,a2}}$ are determined by measuring $|S_{21}(\omega)|$ at a fixed magnetic field being off-resonant with the FMR resonance field as indicated by the vertical dashed violet line in Fig.\,\ref{Fig:1}(b)($l=1$). The result is shown in Fig.\,\ref{Fig:1}(d). By detuning $H_{\mathrm{ext}}$ from the resonance field $H_{\mathrm{res}}$ we ensure that the $\eta_{\mathrm{a1,a2}}$ are not affected by the interaction with the FMR \cite{Schlitz2022}. Figure\,\ref{Fig:1}(d) shows two absorption signatures attributed to two elastic resonances, for which we determine $\eta_{\mathrm{a1}}/(2\pi)\approx 0.23\pm0.01$\,MHz and $\eta_{\mathrm{a2}}/(2\pi)\approx 0.16\pm0.01$\,MHz (HWHM) by fitting the data to two Lorentzian absorption lines.

\begin{figure}[tbh]	
		\centering
		\includegraphics[width=1.0\columnwidth, clip]{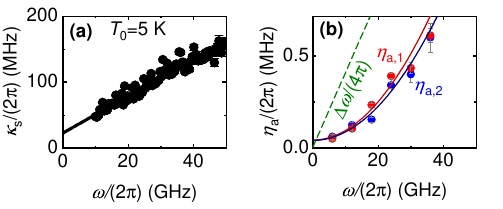}
		\caption{Relaxation rates of the magnetic (a) and elastic (b) subsystems at $T_0=5\,$K as a function of frequency. Lines represent fits to the data with a linear (a) or quadratic (b) frequency dependence for the magnetic and elastic subsystem, respectively. The green dashed line corresponds to half of the frequency splitting of the two transverse elastic modes $\Delta \omega/(4\pi)$ in order to gauge the mode overlap of these two modes.}
		\label{Fig:2}
	\end{figure}

This data analysis allows us to study the loss rates of the magnetic and elastic channels of our CoFe/\ch{Al2O3} sample as function of both frequency and temperature.
In Fig.\,\ref{Fig:2}(a) and (b), we plot the extracted $\kappa_{\mathrm{s}}(\omega)$ and the $\eta_{\mathrm{a},i}(\omega)$ values for $T=\SI{5}{K}$.
We attribute the observed linear dependence in the magnetic relaxation rate $\kappa_{\mathrm{s}}(\omega)$ to viscous Gilbert damping. The corresponding damping parameter $\alpha$ is obtained by fitting the data presented in Fig.\,\ref{Fig:1} to the expression\cite{Nembach2011} $\kappa_{\mathrm{s}}=\kappa_{\mathrm{s}0}+2\alpha \omega$, which yields $\alpha=(2.9\pm0.1)\cdot10^{-3}$. The inhomogeneous linewidth is $\kappa_{\mathrm{s}0}/(2\pi)=(48.0\pm2.4$)\,MHz. Both values are in agreement with room-temperature values reported in Refs.\cite{Flacke2019,Edwards2019}(note that in Fig.\,\ref{Fig:A1} we show, that the magnetization damping of CoFe does not strongly change as function of temperature). 

In analogy to the room-temperature attenuation of elastic modes in GGG in Ref.\,\cite{Schlitz2022, An2023}, we fit the measured elastic damping of the two modes in Fig.\,\ref{Fig:2}(b) to
	\begin{equation}
		\eta_{\mathrm{a},i}(\omega)=\eta_{\mathrm{a},i}^0+\xi_i \omega^2.
		\label{Eq: Standing}
	\end{equation}
The frequency independent relaxation rate $\eta_{\mathrm{a},i}^0$ is assigned to mechanisms related to scattering at defects in the crystal and geometric effects, limiting the phonon confinement to the BAW resonator. The parabolic frequency dependence can be attributed to phonon-phonon scattering, as the phonon density of states scales $\propto \omega^2$ due to the linear dispersion of acoustic phonons for $\hbar\omega/k_\mathrm{B}T<1$, where $\hbar$ is the reduced Planck and $k_\mathrm{B}$ is the Boltzmann constant. We obtain $\eta_{\mathrm{a1,a2}}^0/(2\pi)=(43\pm17)\,$kHz, $(41\pm18)\,$kHz and $\xi_{1/2}/(2\pi)=(1.3\pm0.1)\,\cdot10^{-8}/\mathrm{GHz},\,(1.2\pm0.1)\cdot10^{-8}/\mathrm{GHz}$, which are one order of magnitude lower compared to the values found for GGG in Ref.\cite{Schlitz2022} at room temperature. 

\begin{figure}[tbh]	
		\centering
		\includegraphics[width=1.0\columnwidth, clip]{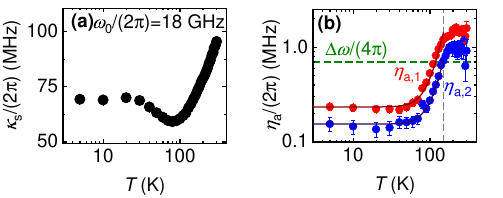}
		\caption{Relaxation rates of the magnetic (a) and elastic (b) subsystems at $\omega/(2\pi)\approx\omega_0/(2\pi)=$18\,GHz as a function of temperature. The continuous lines in panel (b) represent fits to Eq.\,(\ref{Eq: Landau-Rumer}) and the green dashed line indicates half of the frequency splitting of the two transverse elastic modes $\Delta \omega/(4\pi)$. For $\eta_{\mathrm{a},i} > \Delta\omega/2$ the linewidth of the two transverse acoustic modes becomes larger than their frequency spacing. }
		\label{Fig:3-linewidth}
	\end{figure}

The observation of pronounced standing elastic waves in BAW and Fabry-Perot resonators is expected when the effective (acoustic) decay lengths $\delta_{1/2}=v_{\mathrm{t}}/\eta_{\mathrm{a}1/2}(\omega) > 2L$. Given the frequency dependence presented in Fig.\,\ref{Fig:2}\,(b), we thus expect to meet this criterion up to $\omega/(2\pi)\simeq50$\,GHz at $T=5\,$K.
The elastic relaxation rate derived from the fit in Fig.\,\ref{Fig:2}(b) translates into a quality factor of the BAW resonator of $Q\simeq (20-25)\cdot10^3$ at $\omega/(2\pi)=1$\,GHz, which is comparable to the values of high overtone bulk acoustic resonators reported in Refs.\,\cite{Zhang2006, Gokhale2020}. Here, we speculate that the imperfect plane-parallelism of the top and bottom surface and finite surface roughness of the substrate limit the measured $Q$-factor, as such imperfections can cause acoustic loss channels (for details see Ref.\cite{Selfcife2022}).
The parabolic frequency dependence of $\eta_\mathrm{a}$ can also be seen as an asset for the generation of helical phonons, which requires the controlled excitation of a superposition of both orthogonal shear waves by a single chiral drive (the FMR mode). For the investigated sample, we identify $\omega/2 \pi <0.8$\,GHz as the frequency range where this can be achieved from Fig.\,\ref{Fig:2}(b). In this regime, the acoustic relaxation rate exceeds the frequency separation of the two acoustic modes $\Delta \omega/2 < \eta_{\mathrm{a1,a2}} $ (see green dashed line in Fig.\,\ref{Fig:2}(b)). Obviously, in this case the simultaneous excitation of the two orthogonal acoustic modes is possible. Alternatively, the temperature dependence of the elastic relaxation and the sound velocity can also be explored for this purpose as discussed in the context of Fig.\,\ref{Fig:3-linewidth}.
	
The temperature dependence of the off-resonant elastic and magnetic relaxation rates $\kappa_{\mathrm{s}}$ and $\eta_{\mathrm{a},i}$ in the vicinity of $\omega_0/(2\pi)=$18\,GHz is presented in Fig.\,\ref{Fig:3-linewidth}.
The magnetization damping rate $\kappa_{\mathrm{s}}$ is approximately constant up to $T=30\,$K followed by a minimum at $T=100\,$K and a continuous increase for higher $T$. The regime $4\,\mathrm{K}<T<100\,$K matches the behavior of Permalloy in our previous studies \cite{Muller2021a,Suraj2020a} and theoretical predictions for the magnetic damping in $3d$ transition metals\cite{Gilmore2007}. The increase of $\kappa_{\mathrm{s}}$ for higher $T$ can be explained with the fact that the linewidth of the acoustic modes increases with $T$, resulting in a more effective phonon pumping contribution to the magnetization relaxation \cite{Streib2018, Sato2021}(see Appendix \ref{Appendix C}).

We note that the contribution of magnon-magnon scattering to $\kappa_\mathrm{s}$ also increases with $T$, though its contribution to the FMR linewidth has been found to be small for $3d$ transition metals\cite{Zhao2016}. 
	
For the elastic relaxation rates $\eta_{\mathrm{a,i}}$ in Fig.\,\ref{Fig:3-linewidth}(b), we observe nearly constant values from $300\,$K down to $T\approx 180\,$K. Notably, the formation of standing acoustic waves is also detected at elevated temperatures up to room temperature at $\omega_0/(2\pi)=$18\,GHz as $\delta_{1/2}(T=\mathrm{RT})\approx0.6\,\mathrm{cm}/1.0\,\mathrm{cm}\simeq2L$.
For lower temperatures, $\eta_{\mathrm{a1,a2}}$ decrease strongly down to $T=80$\,K and remain roughly constant for lower $T$. The temperature dependence in the low temperature regime is governed by the Landau-Rumer mechanism\cite{Landau1937} as experimentally found in Ref.\cite{Transducers1964}. This model predicts \cite{Transducers1964}
	\begin{equation}
		\eta_{\mathrm{a},i}(T)=\iota_{\mathrm{a},i}^0+\beta_{\mathrm{a},i} T^4.
		\label{Eq: Landau-Rumer}
	\end{equation}
The comparison with our data yields the parameters $\iota_\mathrm{a1,a2}^0/(2\pi)=(0.22\pm0.02)\,\mathrm{MHz},\,(0.16\pm0.01)$\,MHz and
$\beta_\mathrm{a1,a2}/(2\pi)=(0.97\pm0.03)\cdot10^{-3}\mathrm{\,Hz/K^4},(2.28\pm0.16)\cdot10^{-3}\mathrm{\,Hz/K^4}$, which are in reasonable agreement with the values in Ref.\,\cite{Transducers1964} (see Appendix \ref{AppendixB}). Note that $\iota_\mathrm{a1,a2}^0$ can be limited by the imperfect plane-parallelism of the substrate surfaces as well as the surface morphology\cite{Selfcife2022}. For temperatures $T>150$\,K, the phonon modes are highly excited and hence we observe a roughly constant $\eta_\mathrm{a}$\cite{Herring1954}.

The strongly increased $\eta_\mathrm{a}$ at elevated temperatures translates to a more pronounced mode overlap of the two transverse elastic modes, which are split in frequency by $\Delta \omega$. Under these conditions, the driving of the FMR results in a simultaneous excitation of both transverse acoustic modes and the generation of phonons with a defined circular polarization. Specifically, we identify the condition $\Delta \omega/2\leq\eta_\mathrm{a}$ for this process (see green dashed line in Fig.\,\ref{Fig:3-linewidth}(b)), which we find fulfilled for $T\geq\SI{150}{K}$ and $\omega_0/(2\pi)=\SI{18}{GHz}$. This underlines the potential of engineering the relative orientation of the plane-parallel substrate surfaces with respect to the crystallographic directions in order to achieve an efficient generation of phonons with circular polarization.

Next, we focus on the quantification of the coupling rate between the FMR mode and the resonant modes of the BAW resonator. To this end, we model the coupled system following the approach in Refs.\,\cite{An2020, Selfcife2022, An2023} (see Eq.\,(\ref{Eq: Klein2}) and Appendix\,\ref{Appendix D}). Solving for the amplitude of the magnetic excitation $m_{x,y}$ allows one to express the recorded FMR absorption line as a Lorentzian with a probe frequency dependent linewidth $\Delta H_{\mathrm{MEC}}$ corresponding to the loss rate \cite{Herskind2009} (see also Fig.\,\ref{Fig:1}(c))
\begin{equation}
	\tilde{\kappa}_\mathrm{s}\left(\omega\right)=\kappa_\mathrm{s}\left(\omega\right)+\frac{\eta_{\mathrm{a,i}}g_{\mathrm{eff}}^2}{4[(\omega-\omega_n)^2+\eta_{\mathrm{a,i}}^2]
	\label{Eq: coupling-rates}}.
\end{equation}
Here, we convert from magnetic field to frequency using $\omega=\frac{g \mu_\mathrm{B}}{\hbar} \mu_0 H_{\mathrm{ext}}$ using $g=2.079$. 

Equation\,\eqref{Eq: coupling-rates} in combination with the extracted elastic relaxation rates $\eta_{\mathrm{a},i}$ and the undisturbed FMR linewidth $\kappa_\mathrm{s}(\omega)$ allows us to extract the effective coupling rate

$g_\mathrm{eff}$ from the frequency evolution of the FMR linewidth $\Delta H_\mathrm{MEC}$. 

\begin{figure}[tbh]	
		\centering
		\includegraphics[scale=1.0]{ 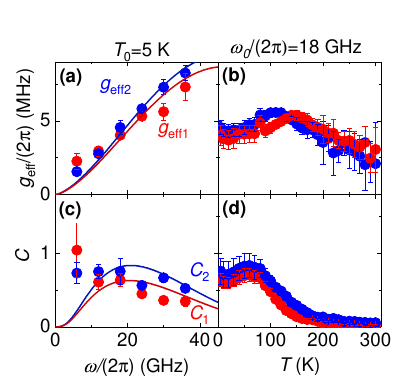}
		\caption{(a)\,Magnetoelastic coupling rate $g_{\mathrm{eff}}$ as a function of $\omega$ (a) and temperature $T$ (b). (c)\,Sample cooperativity $C$ at $\omega_{\mathrm{0}}/(2\pi)=18$\,GHz as a function of $\omega$ (c) and $T$ (d). Continuous lines in panels (a) and (b) represent fitting curves to Eqs.\,(\ref{Eq:Magnetoelastic}) and $C=g_{\mathrm{eff}}^2/(2\eta_{\mathrm{a,i}}\kappa_{\mathrm{s}})$ using the fitting results of Fig.\,\ref{Fig:2}.}
		\label{Fig-4-Coupling-and-cooperativity}
	\end{figure}

Figure\,\ref{Fig-4-Coupling-and-cooperativity}(a) and (b) (see also Appendix\,\ref{Appendix D}) shows $g_{\mathrm{eff}}$ as a function of $\omega$ at $T_0=5$\,K, and as a function of $T$ for $\omega_0/(2\pi)=$18\,GHz.

We observe an increase in $g_{\mathrm{eff}}$ with increasing $\omega$. This behavior is corroborated by the theoretical model presented in Refs.\,\cite{An2020,Schlitz2022,Streib2018},
 which describes the mode overlap between the magnetic and the mechanical mode. In detail, 
		\begin{equation}
			g_{\mathrm{eff}}(\omega)=B\sqrt{\frac{2g\mu_{\mathrm{B}}}{\hbar\omega M_{\mathrm{s}}\tilde{\rho} d L} }\left[1-\mathrm{cos}\left(\omega \frac{d}{\tilde{v}_t}\right)\right],
			\label{Eq:Magnetoelastic}
		\end{equation}
where $M_s$ is the saturation magnetization and $\tilde{\rho}$ is the volume density of the CoFe layer. For a quantitative comparison, we use the following materials and design parameters: $M_\mathrm{s}=M_{\mathrm{eff}}=1.90\cdot10^6$\,A/m, $g=2.079$ \footnote{From Fig.\,\ref{Fig:A1} in Appendix \ref{AppendixA}, we find that $g$ is approximately temperature-independent, whereas $M_\mathrm{s}$ reduces only slightly with increasing temperatures. For the sake of simplicity, we assume a constant $M_{\mathrm{s}}$.}, $d = \SI{30}{nm}$, as well as the the transverse velocity $\tilde{v}_{\mathrm{t}} = \SI{3170}{m/s}$ \cite{Selfcife2022} and volume density $\tilde{\rho} = \SI{8110}{kg/m^3}$ of the CoFe layer at room temperature \cite{Schwienbacher2019}. In addition, we use $L = \SI{510}{\micro m}$. Fitting Eq.\,(\ref{Eq:Magnetoelastic}) to the data presented in Fig.\,\ref{Fig-4-Coupling-and-cooperativity}\,(a) yields
the magnetoelastic coupling constant $B =(13.5\pm0.6)\cdot10^6\,\mathrm{J/m^3}, (12.3\pm0.8)\cdot10^6\,\mathrm{J/m^3}$, which is consistent for the two transversal phonon branches and slightly lower than $B =15.7\cdot10^6\,\mathrm{J/m^3}$, which we estimated for the alloy composition \cite{Selfcife2022}.
As we record a maximum $\eta_{\mathrm{a},i}/(2\pi)<\SI{1}{MHz}<g_\mathrm{eff}/(2\pi)$, we are operating in a regime corresponding to the Purcell enhanced regime in the experimentally investigated frequency range ($5\,\mathrm{GHz}\leq f\leq 50\,\mathrm{GHz}$) at $T=5\,$K \cite{Zhang2014}. In Fig.\,\ref{Fig-4-Coupling-and-cooperativity}(c), we plot the cooperativity $C_{1/2}=g_\mathrm{eff}^2/(2\kappa_\mathrm{s} \eta_\mathrm{a1,a2})$ as function of $\omega$. 

Here, the continuous lines represent the calculated $C$ using our fitted $\eta_{\mathrm{a,i}}$ described by Eq.\,(\ref{Eq: Standing}), $g_{\mathrm{eff}}$ from Eq.\,(\ref{Eq:Magnetoelastic}) and using the parameterized magnetic linewidth evolution $\kappa_{\mathrm{s}}(\omega)$ obtained from Fig.\,\ref{Fig:2}(a). Note that the maximum cooperativity is obtained around 18\,GHz and does not coincide with a maximum in $g_{\mathrm{eff}}$ from Eq.\,(\ref{Eq:Magnetoelastic}) due to the frequency dependence of $\eta_{\mathrm{a},i}$ (see also Ref.\cite{An2020}). We observe a good agreement between theory and experiment for $\omega/(2\pi)>15$\,GHz, but do not observe the predicted reduction in $C$ for lower frequencies. As evident in Fig.\,\ref{Fig-4-Coupling-and-cooperativity}(a), we can attribute this to the larger values of $g_{\mathrm{eff}}$ at low $\omega$ extracted from the experiment as compared to the theoretical expected values. 

In Fig.\,\ref{Fig-4-Coupling-and-cooperativity}(b), we plot the temperature dependence of $g_{\mathrm{eff}}$ at $\SI{18}{GHz}$. Simplistically, we would expect a temperature independent behavior as $g_{\mathrm{eff}}$ originates from the mode overlap. However, we observe a slight increase with increasing $T$, which results in a maximum in $g_{\mathrm{eff}}$ at 150\,K for mode 1 and at 110\,K for mode 2. We attribute this to variations of the acoustic properties of the CoFe thin film and sapphire substrate, which at the peak position results in a minimum mismatch of their acoustic impedances ($Z=\sqrt{\rho v_\mathrm{t}}$) and thereby allows for the more efficient injection of phonons for $Z_{\mathrm{CoFe}}=Z_{\ch{Al2O3}} $\cite{Sato2021}.
The reduction in $g_{\mathrm{eff}}(T)$ for higher temperatures is in agreement with the expected reduction of $B$ for $3d$ transition metals at elevated $T$\cite{Barangi2015}. Comparing the evolution of $g_{\mathrm{eff}}(T)$ with $\eta_{\mathrm{a}}(T)$ in Fig.\,\ref{Fig:3-linewidth}(b), we find $g_{\mathrm{eff}}>\eta_{\mathrm{a}}$ for $T<130$\,K for mode 1 and for $T<160$\,K for mode 2.
For the temperature dependence of the cooperativity in Fig.\,\ref{Fig-4-Coupling-and-cooperativity}(d), we observe a peak at around 60\,K, which can be tracked to the reduced $\kappa_{\mathrm{s}}$ in this temperature range (see Fig.\,\ref{Fig:3-linewidth}(a)), followed by a strong decrease with increasing temperatures, mirroring the increase in the $\eta_{\mathrm{a},i}$ in this temperature range (see Fig.\,\ref{Fig:3-linewidth}(b)). 

\begin{figure}[tbh]	
		\centering
		\includegraphics[width=1.0\columnwidth, clip]{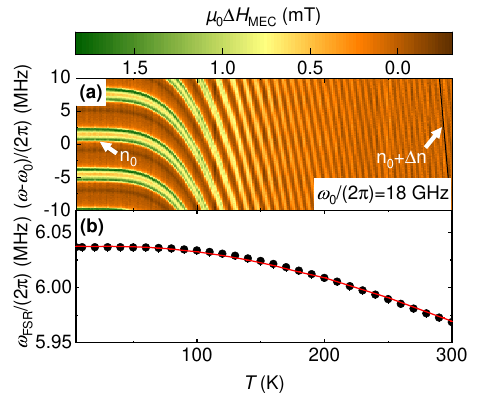}
		\caption{(a)\,Fitted change in FMR linewidth $\Delta H$ as function of $T$ around $\omega_{\mathrm{0}}/(2\pi)=18$\,GHz.\,(b) Free spectral range $\omega_{\mathrm{FSR}}$ as function of $T$. The red continuous line represents a theory curve for the temperature dependence of $\omega_{\mathrm{FSR}}(T)$ using Eq.\,\eqref{eq:modefrequencies}.}
		\label{Fig:5}
	\end{figure}

To map out the temperature evolution of the BAW resonances, we plot the FMR linewidth change $\Delta H_{\mathrm{MEC}}=\Delta H-\Delta H_0$  in Fig.\,\ref{Fig:5} . Here, $\Delta H_0$ is the uncoupled FMR linewidth around $\omega_0/(2\pi)=18$\,GHz, which is  a function of temperature and frequency, and $\Delta H$ is the FMR linewidth obtained via Eq.\,\eqref{formula: s21-real}.  At resonant coupling (see Fig.\,\ref{Fig:1}), the linewidth $\Delta H$ increases by $\Delta H_\mathrm{MEC}$ due to the coupling of the FMR mode to the acoustic mode, appearing as additional dissipation channel. The representation used in Fig.\,\ref{Fig:5} also allows to visualize resonant magnon-phonon coupling up to room temperature. 
We observe that the peaks in $\Delta H_{\mathrm{MEC}}$ broaden and overlap for increasing $T$. In addition, we observe that the peaks shift to lower frequencies above 75\,K. As we do not selectively track one bulk acoustic resonance, we also find higher mode bulk acoustic resonance harmonics within the spectrum. In detail, for the displayed frequency window we count the transit of $\Delta n$=35 bulk acoustic modes when varying temperature between $T=5$\,K and $T=295$\,K. 

In addition to the two prominent MEC features with positive $\Delta H_{\mathrm{MEC}}$, we also find a faint signature with $\Delta H_{\mathrm{MEC}}<0$ for $T<100\,K$. We attribute this feature to the coupling of the Kittel mode to standing waves in the substrate of an unexpected additional phononic branch with a slightly altered propagation velocity compared to the two transverse elastic modes. Seemingly the reduction of the bare FMR linewidth suggests that the elastic mode reduces the FMR damping, potentially due to an mode interference effect. From the evolution of the MEC features of $\Delta H_{\mathrm{MEC}}$ in Fig.\,\ref{Fig:5}(a), we can extract the free spectral range $\omega_{\mathrm{FSR}}(T)$ as shown in Fig.\,\ref{Fig:5}(b) and compare its temperature dependence with the expected behavior of $\omega_{\mathrm{FSR}}(T)$ modeled by
		\begin{equation}
		 \omega_{n,i}=n\pi/[(d/\tilde{v}_{\mathrm{t}}+L(T)/v_\mathrm{t}(T))]. \label{eq:modefrequencies}
		\end{equation} 
As the thickness of the FM layer ($d\approx \SI{30}{nm}$) is thin compared to the substrate ($L\approx \SI{510}{\micro m}$), we expect that the temperature dependence of the acoustic properties is dominated by the substrate properties.
In detail, the transverse velocity is given by $v_{\mathrm{t}}(T)=\sqrt{G(T)/\rho(T)}$, where $G(T)$ and $\rho(T)$ are the shear modulus and the volume density of the substrate. By accounting for the temperature dependence in $\rho(T)$ and $L(T)$ via the thermal expansion coefficient $\alpha^*(T)$ using Eq.\,(6) in Ref.\,\cite{White1997} and the temperature evolution of $G(T)$\cite{Tefft1966}, we can calculate the expected behavior for $\omega_{\mathrm{FSR}}(T)$ (solid line in Fig.\,\ref{Fig:5}(b)). The result corroborates the experimental observations. Notably, by monitoring the temperature dependence of the high-overtone BAW resonances, we are able to observe changes in $\omega_{\mathrm{FSR}}/(2\pi)$ down to the kHz-range, which via Eq.\,\eqref{eq:modefrequencies} and using the parameters $L\simeq 510\,\mu$m and $v_{\mathrm{t}}\simeq6.1$\,km/s translates to a relative change in substrate thickness of $\Delta L/L\approx 1.7\cdot 10^{-4}$. This value is comparable to the sensitivity of high precision strain sensors\cite{Kim2022,Wang2021c}.

\section{Conclusion and Outlook}
In summary, we study magnetoelastic coupling phenomena in CoFe/crystalline sapphire hybrid samples. In particular, we extract the frequency and temperature dependence of both the magnetic and elastic loss rates $\kappa_{{\mathrm{s}}}$ and $\eta_{\mathrm{a}}$ as well as the effective magnetoelastic coupling rate $g_{\mathrm{eff}}$ and cooperativity $C$ from broadband magnetic resonance experiments. We also identify optimal working points for the realization of magnetoelastic devices. Here, in particular the strong increase in the elastic loss rates $\eta_{\mathrm{a}}$ with increasing $T$ demonstrates the advantage of performing these experiments at cryogenic temperatures to push coupled magnon-phonon systems to stronger coupling regimes such as the high cooperativity Purcell-enhanced regime\cite{Zhang2014}. From a different perspective, the strong increase of $\eta_{\mathrm{a}}$ at elevated temperatures gives rise to a substantial mode overlap of the two transverse elastic waves, which is a prerequisite for the generation of circularly polarized phonons in the non-magnetic crystal. Hence, in particular the sample temperature can be viewed as a control knob for the controlled pumping of either circular or linear phonons. Regarding the magnetic loss rate, we find an almost temperature independent Gilbert damping, which renders the CoFe/\ch{Al2O3} material platform superior to the conventionally used YIG/GGG system\cite{An2020,An2022, An2023, Schlitz2022} regarding their performance at cryogenic temperatures. Finally, the evolution of the free spectral range $\omega_{\mathrm{FSR}}$ allows to study the temperature dependence of the thermal expansion coefficient and shear modulus with high sensitivity. This allows one to use CoFe/BAW resonators for the realization of magnetic micro-actuators and systems (MAGMAS)\cite{Ueno2002, Qhobosheane2020,Cugat2003,Reyne2005}.

	\section*{Acknowledgments}
	We acknowledge financial support by the Deutsche Forschungsgemeinschaft (DFG, German Research Foundation) via Germany’s Excellence Strategy EXC-2111-390814868 and TRR 360 (Project-ID 492547816). This research is part of the Munich Quantum Valley, which is supported by the Bavarian state government with funds from the Hightech Agenda Bayern Plus.

\appendix
\section{Analysis of the ferromagnetic resonance lineshape}
\label{Appendix: fitting function}
The Polder susceptibility of the ferromagnetic resonance for a static magnetic field applied along the surface normal of a magnetic thin film, i.e. the out-of-plane (oop) geometry, is given by \cite{Sato2021, Polder1949, Muller2021a}
\begin{widetext}
\begin{equation}
	\begin{aligned}
		\hat{\chi}_{\mathrm{P}}=\begin{pmatrix}
			\chi_{{\mathrm{xx}}} & \chi_{{\mathrm{xy}}}\\
			\chi_{{\mathrm{yx}}} & \chi_{{\mathrm{yy}}}
		\end{pmatrix}=\frac{\mu_0 M_\mathrm{s}}{\Delta(\omega)}\begin{pmatrix}
			\gamma \mu_0 (H_{\mathrm{ext}}- M_\mathrm{s})-i\kappa_\mathrm{s} & -i\omega\\ i\omega & \gamma \mu_0 (H_{\mathrm{ext}}- M_\mathrm{s})-i\kappa_\mathrm{s} 
		\end{pmatrix}
	\end{aligned}
	\label{polderold}
\end{equation}

\end{widetext}
assuming a negligible uniaxial anisotropy constant $K_{\mathrm{u}}=0$. 
Here, $\Delta(\omega)$ is the determinant of $\hat{\chi}_{\mathrm{P}}$, $M_\mathrm{s}$ is the saturation magnetization, $H_\mathrm{ext}$ is the externally applied static magnetic field, $\kappa_\mathrm{s}$ is the FMR linewidth, and $\mu_0$ is the magnetic vacuum permeability. We fit our experimental spectroscopy data $S_{21}(H_{\mathrm{ext}})|_{\mathrm{\omega}}$ shown exemplary in Fig.\,\ref{Fig:1}(c) using
\begin{equation}
S_{21}(H_{\mathrm{ext}})|_{\mathrm{\omega}}=C_0+C_1\cdot H_{\mathrm{ext}}-iA e^{i\phi}\cdot\frac{\chi_{\mathrm{yy}}(H_{\mathrm{ext}})}{\mu_0 M_{\mathrm{s}}}.
\label{formula: s21-real}
\end{equation}
To account for the complex microwave transmission background $S_{21}^0$ in our experiments, we include a linear function of the form $S_{21}^0=C_0+C_1 H_{\mathrm{ext}}$ with the complex offset $C_0$ and slope $C_1$ following the approach used in Refs.\,\cite{Nembach2011, Muller2021a}. Furthermore, in Eq.\,\eqref{formula: s21-real}, $A$ is an amplitude and $\phi$ is a phase factor describing the microwave detection circuit. 

\section{Temperature dependence of the magnetic resonance parameters of CoFe}
\label{AppendixA}
Figure\,\ref{Fig:A1} summarizes the temperature dependence of parameters describing the magnetization dynamics of the CoFe thin film, which is deposited on a \ch{Al2O3} substrate and investigated in the main part of the manuscript. The $g$-factor displayed in Fig.\,\ref{Fig:A1}\,(a) shows only a minimal decrease with increasing $T$. The effective magnetization $M_{\mathrm{eff}}$ presented in Fig.\,\ref{Fig:A1}\,(b) decreases for increasing temperature due to the thermal excitation of magnons. To asses the magnetization damping properties, we analyze the frequency dependence of the FMR linewidth using \cite{Nembach2011}
\begin{equation}
\kappa_\mathrm{s} =\kappa_\mathrm{s0} + 2 \alpha \omega .
\end{equation}
Here, $\kappa_\mathrm{s0}$ represents the inhomogeneous linewidth and $\alpha$ is the Gilbert damping parameter. We note that the Gilbert damping parameter $\alpha$ (see Fig.\,\ref{Fig:A1}\,(c)) is approximately temperature independent, while we observe a slight increase in the inhomogeneous linewidth $H_{\mathrm{inh}}=\frac{\hbar \kappa_\mathrm{s0}}{g \mu_0 \mu_\mathrm{B}}$ for increasing temperatures (see Fig.\,\ref{Fig:A1}(d)), which we attribute to nonlinear damping induced by thermal magnons.
		\renewcommand{\thefigure}{B1}
		
		\begin{figure}[htbp]	
			\centering
			\includegraphics[width=1.0\columnwidth, clip]{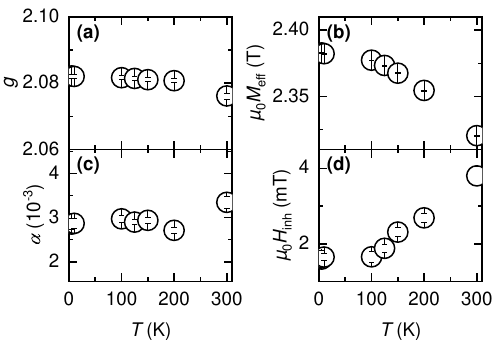}
			\caption{Temperature dependence of the parameters in the CoFe thin film deposited on an \ch{Al2O3} substrate.\,(a) As expected, the $g$-factor is about independent of $T$.\,(b)\,The effective magnetization $M_{\mathrm{eff}}$ decreases with increasing temperature due to the thermal excitation of magnons.\,(c)\,The Gilbert damping $\alpha$ is approximately constant as function of temperatures.\,(d)\,Inhomogeneous linewidth broadening $H_{\mathrm{inh}}$ as function of $T$. We observe a slight increase in $H_{\mathrm{inh}}$ with increasing $T$, which we attribute to nonlinear damping induced by thermal magnons.}
			\label{Fig:A1}
		\end{figure}
		
\section{Temperature dependence of the elastic damping }
\label{AppendixB}
		For the Landau-Rumer mechanism, the attenuation coefficient $\nu(T)$ of the transverse elastic waves is given by\cite{Transducers1964}
		\begin{equation}
			\nu(T)=60\gamma_{\mathrm{G}}^2\frac{ k_{\mathrm{B}} }{Mv_{\mathrm{t}}^3\Theta_{\mathrm{D}}^3}\cdot \omega\cdot T^4,
			\label{Eq: beta}
		\end{equation}
		where $\Theta_{\mathrm{D}}$ is the Debye temperature and $\gamma_{\mathrm{G}}$ is the Grüneisen parameter, $M$ is the average atomic mass. Using the attenuation coefficient $\nu(T)$ in combination with the characteristic decay length $\delta_\mathrm{LR}=v_{\mathrm{t}}/\eta_{\mathrm{a}}$\cite{An2020}, defined as the distance, at which the amplitude of the phonon mode has decayed to $1/e$ of its initial value, we can determine the elastic relaxation rate by using 
		\begin{equation}
		\frac{|u(\delta_\mathrm{LR})| }{|u(0)|}=\mathrm{exp}(-\nu\delta_\mathrm{LR})=\mathrm{exp}(-1).
			\label{Eq: etaA}
		\end{equation}
	 Here, $|u(\mathrm{z'})|$ is the magnitude of acoustic modes after propagating distance $\mathrm{z'}$ through the \ch{Al2O3} substrate in the $z$-direction. From Eq.\,\eqref{Eq: etaA}, we obtain the relation $\eta_{\mathrm{a}}=\nu v_{\mathrm{t}}=\beta_{\mathrm{a}}T^4$. By inserting $\gamma_{\mathrm{G}}=2$, $(Mv_{\mathrm{t}}^2/k_\mathrm{B})=135000$\,K and $\Theta_{\mathrm{D}}=1000$\,K from Ref.\,\cite{Transducers1964} and $v_{\mathrm{t}}=6.17$\,km/s \cite{Selfcife2022} in Eq.\,(\ref{Eq: beta}), we obtain $\beta_{\mathrm{a}}/(2\pi)\approx 3.2\cdot10^{-2}\,\mathrm{Hz/K^4}$, which is in reasonable agreement with the obtained fitting values $\beta_\mathrm{a1,a2}/(2\pi)=(0.97\pm0.03)\cdot10^{-3}\mathrm{\,Hz/K^4},(2.28\pm0.16)\cdot10^{-3}\mathrm{\,Hz/K^4}$ in the main text.

\section{Determination of the effective coupling rate from the frequency dependent FMR linewidth}
\label{Appendix D}
When the FMR mode couples resonantly to an acoustic mode of the BAW resonator, the resulting linewidth modification $\Delta H_{\mathrm{MEC}}$ allows one to quantify the magnon-phonon coupling strength $g_{\mathrm{eff}}$. Here, we derive an explicit expression for the FMR linewidth evolution as function of the probe frequency $\omega$. To this end, we use the same formalism as in Ref.\,\cite{An2020} and describe our FM/BAW resonator system as interacting harmonic resonators. In particular, we only consider one individual elastic BAW mode coupling to the magnetic Kittel mode. The differential equations describing this model read as
\begin{align}
	\begin{aligned}
		( \omega- \omega_{\mathrm{mag}}+i\kappa_\mathrm{s})m^\mathrm{x,y}&=g_{\mathrm{eff}}u_{n}^\mathrm{x,y}/2+\zeta h^\mathrm{x,y}\\
		(\omega-\omega_{n}+i\eta_{\mathrm{a}})u_{n}^\mathrm{x,y}&=g_{\mathrm{eff}} m^\mathrm{x,y}/2.
	\end{aligned}
	\label{Eq: Klein2}
\end{align}
Here, $\omega$ is the microwave probe frequency, $\omega_{\mathrm{mag}}$ is the magnetic resonance frequency, $\omega_n$ is the resonance frequency of the elastic mode, $\eta_{\mathrm{a}}$ ($\kappa_\mathrm{s}$) is the damping rate of the elastic (magnetic) system, and $u^\mathrm{x,y}$ ($m^\mathrm{x,y}$) are the amplitudes of the elastic (magnetic) excitations. Note that $\omega_n$ is connected to the phonon propagation velocities via Eq.\,(\ref{eq:modefrequencies}). The parameter $\zeta$ describes the inductive coupling to the antenna and $h^\mathrm{x}$ and $h^\mathrm{y}$ are the in and out-of-phase amplitude of the microwave magnetic field generated by the coplanar waveguide. We solve Eq.\,(\ref{Eq: Klein2}) for $m^\mathrm{x,y}$ and obtain
\begin{align}
	m^{\mathrm{x,y}}=-\zeta h^\mathrm{x,y}\frac{(\omega-\omega_{n}+i\eta_\mathrm{a})}{(\omega-\omega_n+i\eta_\mathrm{a})(\omega-\omega_{\mathrm{mag}}+i\kappa_\mathrm{s})+g_{\mathrm{eff}}^2/4}.
\end{align}
We can rewrite this expression as
\begin{widetext}
\begin{align}
	m^{\mathrm{x,y}}=-\frac{\zeta h^\mathrm{x,y}}{(\omega-\omega_{\mathrm{mag}})-g_{\mathrm{eff}}^2/(4C(\omega))(\omega-\omega_n)+i\cdot[\kappa_\mathrm{s}+g_{\mathrm{eff}}^2\eta_{\mathrm{a}}/(4C(\omega))]},
	\label{Eq: mxy2}
\end{align}
\end{widetext}
where we have defined $C(\omega)=(\omega-\omega_n)^2+\eta_{\mathrm{a}}^2$. The imaginary part in the denominator of Eq.\,\eqref{Eq: mxy2} can be interpreted as a modified magnetic relaxation rate $\tilde{\kappa}_\mathrm{s}$ in the presence of magnetoelastic coupling and is given by
\begin{equation}
	\tilde{\kappa}_\mathrm{s}=\kappa_\mathrm{s}+\frac{\eta_{\mathrm{a}}g_{\mathrm{eff}}^2}{4[(\omega-\omega_n)^2+\eta_{\mathrm{a}}^2]}
\end{equation}
In resonance with the phononic modes, we obtain
\begin{equation}
	\tilde{\kappa}_\mathrm{s}(\omega=\omega_n)=\kappa_\mathrm{s}+\frac{g_{\mathrm{eff}}^2}{4\eta_{\mathrm{a}}}
	\label{Eq: kappas2}
\end{equation}
Alternatively, 
\begin{align}
\begin{aligned}
	g_{\mathrm{eff}}&=2\sqrt{\eta_\mathrm{a}(\tilde{\kappa}_\mathrm{s}(\omega=\omega_n)-\kappa_\mathrm{s})}=\\ &=\sqrt{2\eta_\mathrm{a}\gamma[\mu_0(\Delta H(\omega=\omega_n)-\Delta H_0)]}.
	\label{coupling2}
	\end{aligned}
\end{align}
Hence, in our experiments, the effective coupling rate between the Kittel mode and the elastic modes of the acoustic resonator can be determined from the resonant change of the FMR linewidth.

In Fig.\,\ref{Fig:S3}, we plot the linewidth modulation of the FMR linewidth $\Delta H$ as function of microwave frequency around 18\,GHz at $T=5$\,K. The continuous red line represents the calculated linewidth modulation following Eq.\,\eqref{Eq: coupling-rates} together with the extracted values for $\kappa_\mathrm{s}$, $\eta_\mathrm{a1,a2}$ and $g_{\mathrm{eff1/2}}$ from the main text and plotting the sum of the two individual resonances with resonance frequencies at $\omega_{n1}/(2\pi)=18.0008$\,GHz and $\omega_{n2}/(2\pi)=18.0024$\,GHz for the two transverse elastic modes.
\renewcommand{\thefigure}{D1}
\begin{figure}[htbp]	
	\centering
	\includegraphics[width=1.0\columnwidth, clip]{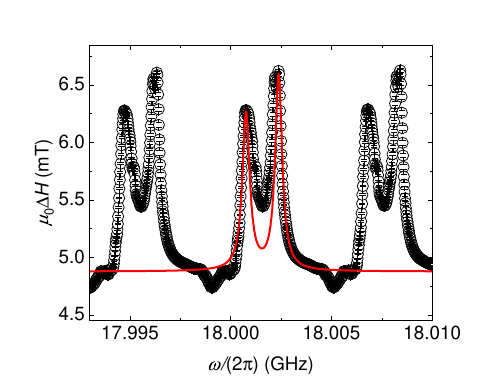}
	\caption{FMR linewidth $\Delta H$ of a CoFe film deposited on a \ch{Al2O3} substrate plotted as a function of $\omega$ around $\omega_0/(2\pi)$=18\,GHz at $T=5$\,K. The continuous red line indicates a plot of the linewidth modulation following Eq.\,\eqref{Eq: coupling-rates} and using the extracted values for $\kappa_\mathrm{s}$, $\eta_\mathrm{a1,a2}$ and $g_{\mathrm{eff1/2}}$ using the sum of the two individual resonances with $\omega_{n1}/(2\pi)=18.0008$\,GHz and $\omega_{n2}/(2\pi)=18.0024$\,GHz. }
	\label{Fig:S3}
\end{figure}
We observe a good quantitative agreement between Eq.\,\eqref{Eq: coupling-rates} and our experimental data. Discrepancies are apparent in particular in the frequency range between the two resonances, which indicates that a more sophisticated model, which accounts for the interaction of the two transverse modes, is required to properly describe the modulation of $\Delta H(\omega)$ in this frequency range.
\section{Phonon pumping contribution to the FMR linewidth}
		\renewcommand{\thefigure}{E1}
\label{Appendix C}
Using Eq.\,\eqref{Eq: coupling-rates}, we study the temperature dependence of the impact of magnetoelastic coupling on the FMR linewidth $\Delta H$.
As we couple to two phonon modes in our experiment, we define the net $\tilde{\kappa}_\mathrm{s}(\omega)$ in our experiment as
\begin{equation}
\tilde{\kappa}_\mathrm{s}(\omega)=\tilde{\kappa}_\mathrm{s1}(\omega,\eta_{\mathrm{a/1}},v_{\mathrm{ft}})+\tilde{\kappa}_\mathrm{s2}(\omega,\eta_{\mathrm{a/2}},v_{\mathrm{st}}).
\label{Eq:kappa12}
\end{equation}

\begin{figure}[htbp]	
	\centering
	\includegraphics[width=1.0\columnwidth, clip]{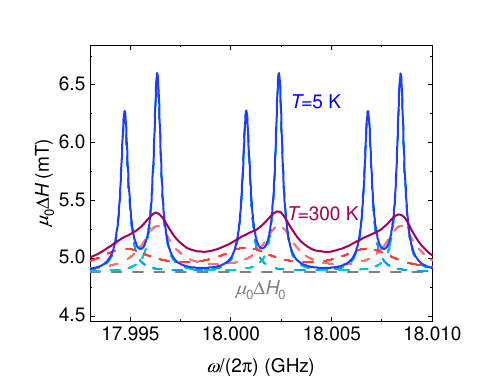}
	\caption{Theory curve for the FMR linewidth using Eq.\,\eqref{Eq:kappa12} as function of $\omega$ around $\omega/(2\pi)=18$\,GHz for cryogenic temperatures (blue line) and room-temperature (red line).Periodic double-peak features are observed. Dotted blue and red lines indicate the individual contributions of $\tilde{\kappa}_{s1}$ and $\tilde{\kappa}_{s2}$ at 5\,K and 300\,K. }
	\label{Fig: S2}
\end{figure}

The resulting impact on the FMR linewidth $\Delta H$ for $T=5$\,K and $T=300$\,K, when using the extracted $\eta_{\mathrm{a,i}}(\omega)$ for the respective temperature from Fig.\,\ref{Fig:3-linewidth}(b), is plotted in Fig.\,\ref{Fig: S2}.

In Fig.\,\ref{Fig: S2}, we observe periodic double-peak features induced from the coupling of the FMR with the two elastic resonances, which are particularly pronounced for the $T=5$\,K case due to the small linewidth of the elastic resonances. For $T=300\,$K, the resonance peaks broaden and overlap. Notably, due to their strongly increased elastic damping, $\tilde{\kappa}_\mathrm{s}$ takes non-zero values even for off-resonant $\omega$-values. Consequently, the net FMR linewidth $\Delta H$ increases for elevated temperatures.
	
	\bibliography{library}

\end{document}